\documentclass{article} 
     \usepackage{amsmath}
\usepackage[super,sort&compress,comma]{natbib} 
     \usepackage{amsfonts}
     \usepackage{amssymb}
     \usepackage{makeidx}
     \usepackage{graphicx}
     \author{Suleiman M Baraka}
   \usepackage{hyperref}
     \usepackage{amssymb}
     \usepackage{amsmath}
     \usepackage{enumerate}
     \usepackage{graphics}
     \usepackage{graphicx}
     \usepackage{url}
     \usepackage{array,arydshln}
     \usepackage{longtable}
     \usepackage{expl3}
     \usepackage[table]{xcolor}
     \usepackage{booktabs,dcolumn}                     % Tables.
     \usepackage[figureposition=bottom]{caption}       % Float captions.
     \usepackage{color}
     \usepackage{cleveref}
     \usepackage[a4paper]{geometry}
     \usepackage[table]{xcolor}
     \usepackage{multirow}
     \usepackage{multicol}
     \usepackage{jabbrv}
     \usepackage{paralist}
     \usepackage{setspace}
\begin{document}
     	
     	\title{Large Scale Earth's Bow Shock with Northern IMF as simulated by \textrm{PIC} code in parallel with \textrm{MHD} model}
     	\author{Suleiman M Baraka$^{(1,2,3)}$}
     	
     	\maketitle
     	
     	$^{(1)}$ Faculty in residence, National Institute of Aerospace,\\
     	100, Exploration Way Hampton,VA\\
     	{suleiman.baraka@nianet.org}
     	
     	$^{(2)}$	Center for Astronomy and Space Sciences-CASSR.\\ Al Aqsa University,P.O.Box 4051, Gaza, Palestine.\\ 
     	{suleiman.baraka@gmail.com}

     	$^{(3)}$ Institut d'Astrophysique de Paris \\98 bis Blvd Arago,75015, Paris, France\\ {baraka@iap.fr}

     	\clearpage
     	\doublespacing
     %%%%%%%%%%%%
     	%%%%%%%%Results%%%%%%%%%%%%%%%%%%
     	%%%%%%%%%%%%%%%%%%%%%%%%%%%%%%%%%%%
     	\begin{abstract}
     		In this paper, we propose a 3D kinetic model  (Particle-in-Cell \textrm{PIC}) for the description of the large scale Earth's bow shock. The proposed version is stable and does not require huge or extensive computer resources. Because  \textrm{PIC} simulations work with scaled plasma and field parameters, we also propose to validate our code by comparing its results with the available   \textrm{MHD} simulations under same  scaled Solar wind (\textrm{SW}) and (\textrm{IMF}) conditions. We report new results from the two models. In both codes the Earth's bow shock position is found to be  $\approx 14.8  R_{E} $ along the Sun-Earth line, and $\approx 29  R_{E} $ on the dusk side. Those findings are consistent with past \textrm{in situ} observations. Both simulations reproduce the theoretical jump conditions at the shock. However, the \textrm{PIC} code density and temperature distributions are inflated and slightly shifted sunward when compared to the \textrm{MHD} results.   Kinetic electron motions and reflected ions upstream  may cause this sunward shift. Species 
     		distributions in the foreshock region are depicted within the transition of the shock (measured  $  \approx  $ 2  $ c/\omega_{pi} $ for $ \Theta_{Bn}=90^{o}$ and $M_{MS}=4.7 $) and in the downstream. The size of the foot jump in the magnetic field at the shock is measured to be  ($ 1.7 c/ \omega_{pi} $). In the foreshocked region, the thermal velocity is found equal to \textrm{213} $ \mathrm{km.sec^{-1}}  $  at $ 15 \mathrm{R_{E}} $ and is equal to \textrm{63}  $\mathrm{km.sec^{-1}}  $  at \textrm{12} $ \mathrm{R_{E}} $ (Magnetosheath region). 
     		Despite the large cell size of the current version of the \textrm{PIC} code, it is powerful to retain macrostructure of planets magnetospheres in very short time, thus it can be used for a pedagogical test purposes. It is also likely complementary  with \textrm{MHD}  to deepen our understanding of the large scale magnetosphere
     	\end{abstract} \clearpage
     	\section{Introduction}
     	Shocks in astrophysical systems are mainly non-relativistic shocks (relativistic shocks are not in the reach of man-made spacecraft). They have widths of order of the ion inertial length ($c/\omega_{pi}$) or ion gyro-radius ($v_{\perp}/\omega_{ci}$ i.e.\  resistive scale $\sim$ $10^{-6}$ mean free path). The collisionless astrophysical shocks is important to understand their effects in dissipating flow-energy, in heating matter, in accelerating particles to high presumably cosmic-ray energies, and in generating detectable radiation from radio to X-rays. \cite{Bykov2011,Treumann2009}.

     	The Earth's bow shock was proposed by Axford\cite{Axford1962} and \cite{Kellogg1962}, since then many theoretical and statistical studies based on space observations have been conducted to study its position and shape for a large set of upstream solar wind plasma and field conditions \cite{Dmitriev2003,Keika2009,Maynard2011,Fontaine2015,Jelinek2010,Mann2008,Meziane2015,Petrukovich2015}. On the other hand\,, there are many  approaches to study the bow shock location, dynamics and physical properties, such as \textrm{Hybrid} models \cite{Omidi2013,Rojas-Castillo2013,Ellison1993}\,,\textrm{MHD} models \cite{Samsonov2007,Shaikhislamov2011,Kowal2011,Welling2013}\,, and  \textrm{PIC} models (\cite{Baraka2007b,Savoini2013, Vapirev2013,Schreiner2014}) and the references therein. 
     	\vskip 0.25cm 
     	\citet{Leboeuf1978a} was the first to use \textrm{MHD} modeling of the global interaction of the magnetosphere with the solar wind. Over the years  these models have increased greatly in their sophistication and scope \cite{Gombosi2000}.  
     	The \textrm{MHD} use only ensemble-averaged parameters which  assume the distribution of the particles velocity  as  a collection of several Maxwellian functions  as in \cite{Winglee2005}. Under the 
     	influence of the magnetic field where velocity distributions along and across the field  lines are generally
     	different ,these calculations do not determine the plasma microphysics \cite{Paschmann1981,Bonifazi1981,Meziane2007,Seki2009,Kronberg2011}. On the other hand\,, the ideal \textrm{MHD} theory may removes the capability for the plasma to act electromagnetically. This restriction severely limits the kind of physics one can do with ideal fluid \cite{Parks2004}. 
     	\vskip 0.25cm 
     	Our code (modified from \citet{Buneman1992}) is a  \textrm{PIC} code.
     	Global \textrm{PIC EM} code has severe constraints on spatial and temporal scales despite it contained more physics than explicitly assuming Ohm's law. The most limiting of them are  $ \Delta x<\lambda _ { De}  $\,, $ c\Delta t<\Delta x/\sqrt { n } $  and   $ { \omega  }_{ pe }\Delta t<2 $ \,, where $\Delta x$ is the grid size, $\Delta t$ is the time step and $\omega_{pe}$ is the electron plasma frequency. However, this method is superior to   \textrm{MHD} simulation in some aspects such as in modeling kinetic processes that separate the electrons and ions dynamics \cite{Nishikawa1997},\cite{Wodnicka2009}, \cite{Cai2006}. 
     	For instance,   \textrm{MHD} has no fundamental length scale in contrast with   \textrm{PIC} simulations for which a gyro-radius can be derived for particles despite the limitation on the $\frac{mi}{me}$  mass ratio.

     	In this paper,a Particle-In-Cell \textrm{PIC} is used for the description of Earth's bow shock. The proposed version is stable and does not require huge or expensive computer resources since we are interested in the large scales of the system ($ 1 R_{E} $) \cite{Baraka2011}. The scaled plasmas and fields parameters used in \textrm{PIC}, was also used to validate our code with available \textrm{MHD} simulations. 
     	
     	\vskip 0.25cm
     	
     	\vskip 0.25cm
     \section{Simulation Models}\label{sec:2}
     	In this section a brief introduction of \textrm{PIC-EM} and \textrm{MHD-GUMICS} models is presented. As in our previous work \cite{Baraka2011}\,, the current version of the code is capable to form the  macrostructure of the  Earth's magnetosphere. The \textrm{MHD} model is introduced based on the \textrm{CCMC} requested run  (\url{http://ccmc.gsfc.nasa.gov/results/viewrun.php?domain=GM&runnumber=Suleiman_Baraka_112610_2}). GUMICS-v4 details are also available at \citet{Janhunen2012}.
     	\subsection{ \textrm{PIC} EM Relativistic Global Code}\label{sec:2.1}
     	\noindent 
     	In our simulation, we use the same initial conditions in \cite{Buneman1980,Buneman1992,Buneman1993,Buneman1995}
     	to generate the macrostructure of  the magnetosphere. The radiating boundary conditions is adopted as in   \cite{Lindman1975} and for the charge description inside the box we used the charge-conserving formulas reported by 
     	\cite{Villasenor1992}. The same initial and boundary conditions were also used  in our previous work \cite{Baraka2007b,Baraka2014,Baraka2013,Baraka2011,Ben-Jaffel2014,Ben-Jaffel2013d}. The grid size in the simulation  should take into account the nonphysical instabilities. In our simulation,  they are taken care of by Courant Condition ($ \delta x, \delta y,\delta z $ \textgreater $ c\delta t $) \,, which satisfies the  inequality  $ \frac{\lambda_{Dei}}{\delta x} $ \textgreater $ \frac{1}{\pi} $ \cite{Birdsall2005}. 
     	
     	\citet{Pritchett2000} thoroughly discussed cons and pros of formulating \textrm{PIC} codes for space simulation. Whilst \citet{Parks2004} clearly stated that understanding collisionless plasma dynamics  requires self-consistent Particle-In-Cell kinetic modeling.
     	The spatial  dimensions of the 3D EM global code used in this simulation is set such that \textrm{OX} is pointing from Earth to Sun, \textrm{OY}  toward  dusk direction and \textrm{OZ} toward north direction. 
     	The dimension of the simulation box is taken equal to ($ 155\Delta \times 105\Delta \times 105\Delta $ )\,, 
     	where the grid size $ \Delta = \Delta x=\Delta y=\Delta z=1 R_{E} $ and  $ \Delta t $ is the time step ($ \omega_{pe} \Delta t=0.22) $.  The simulation box is uniformly  filled up by  2 $ \times 10^{6}$\  of equal  electrons-ions pairs\,, this number is equivalent to a uniform particle density of  $n_{e}=n_{i}= \frac{N}{\Delta^{3}} =0.8 $ pairs per cell.

     	The physical  parameters (normalized) used in our simulation as pairs of numbers (unit-less values for electrons and  ions) are as follow\,, the gyro-frequencies are $ \tilde{\omega}_{ce,i}=\omega_{ce,i} \Delta t=(0.2,0.0125)  $\,, the thermal velocities for the two species are $ {\tilde {v}_{the,i}}=v_{the,i}/(\Delta/\Delta t) =(0.1,0.025)=(B\Delta m_{e}/\Delta t m_{e,i})$, the Debye length is $ \lambda_{De,i}=\tilde {v}_{the,i}/\tilde{\omega}_{pe,i} =(0.11,0.11)$ \,, Larmour gyro-radii are $ \tilde{\rho}_{ce,i}=\tilde {v}_{the,i}/\tilde{\omega}_{ce,i}=(1.25,20) $\,, inertial lengths are $ \tilde{\lambda}_{e.i}=\tilde{c}/\tilde{\omega}_{pe,i}=(0.559,2.236) $. The impinged drift velocity of the solar wind along the Sun-Earth line is $ V_{sw}=-0.25=0.5\tilde{c} $, where the speed of light's normalized value is taken $ \tilde{c}=0.5 $\,,  the ion to mass ratio is $ m_{i} /m_{e}=16 $. The Normalized magnetic field is  $  \tilde{B}=\vec{B}(\frac{q(\Delta t)^{2}}{m_{e}\Delta}) $ , the IMF is northward $ B_{z}(x)=0.2 $\,, the  $  \beta_{e,i}=(1.6,6.4) $. The normalized ion temperature is $ \tilde{T}_{i}=\tilde { v }_{ thi }^{ 2 } m_{i} =0.04$\,, and for electrons the temperature is $ \tilde{T}_{e}=\tilde { v }_{ the }^{ 2 } m_{e} =0.01$ , where the $ "e" $ and the $ "i"  $ denotes electrons and ions respectively .
     	On the other hand\,, Our code was run until it reached a steady state  at  time step $\mathrm{900\Delta t}$, where $\mathrm{\Delta t}$ is the numerical time step \cite{Baraka2011,Baraka2007}.  Moreover in the   \textrm{PIC} simulation, the macroscopic bulk properties of the flow are, $ V_{A}=0.027$\,, $\mathrm{\frac{V_{A}}{V_{SW}}}$=$0.11$, $\mathrm{M_{A}}$=$ 9.219$,$ M_{S}=2.858 $\,, $ \mathrm{M_{MS}}$=$2.730 $,  Plasma parameters were then derived and scaled so that the flow input conditions are used to simulate the same case study by   \textrm{MHD} model. 
     	\vskip 0.25cm
     	
     	\subsection{\bf\large   \textrm{MHD} model: GUMICS }\label{sec:2.2}
     	The Community Coordinated Modeling Center (CCMC) is a multi-agency partnership. The CCMC provides, to the international research community, access to modern space science simulations. In addition, the CCMC supports the transition to space weather operations of modern space research models. More information  about CCMC can be found here ( \url{http://ccmc.gsfc.nasa.gov/})
     	
     	GUMICS is a global solar wind-magnetosphere-ionosphere coupling model. Its solar wind and magnetospheric part is based on solving the ideal   \textrm{MHD} equations and its ionosphere part is based on solving the electrostatic current continuity equation. Advanced numerical methods such as automatically refined Cartesian octogrid and temporal sub-cycling are used to speed up the computation. The computational box dimension is taken  from -224 to +32 $ R_{E} $ ~in~$ GSE~ X $~and from -64 to +64 $ R_{E} $  in $Y$ and $Z$ [\cite{Kullen2004, Palmroth2005, Palmroth2002}, official website is here (\url{http://ccmc.gsfc.nasa.gov/models/modelinfo.php?model=GUMICS}).

     	The inflow boundary conditions are carried out in 5 hours, and the dipole tilt in GSE coordinates  is taken to be zero. The initial solar wind velocity is $  V_{sw}(x) =-500 \mathrm{ km . sec^{-1}}$ \,, the solar wind density is $ \rho_{sw}=5.0 \mathrm{N.cm^{-3}} $. The solar wind temperature is $  T_{e,i}=6.7\times 10^{5}  \mathrm{Kelvin}$. The initial  \textrm{IMF} value in the  \textrm{MHD} code was  $ B_{z}=6.5 \mathrm{nT} $ northward oriented.  The top level (in terms of hierarchy ) of the simulation box has a base grid of $ { (8{ R }_{ E }) }^{ 3 } $. Each cell is broken to 8 sub cells if the refinement exceed a certain limit. The grid size in the magnetohydrodynamics code is changing with the dynamics of the hierarchically adaptive and can reach up to 0.25 $ R_{E} $ \cite{Janhunen2012}.
     	\newpage  
     	%%%%%%%%%%%%%%%%%%%%%%%

	%%%%%%%%%%%%%%%%%%%%%%%%%%%%%%%%%%%
     	\section{Results}\label{secres}
\begin{figure}[htp]
     		\centering

     		\begin{tabular}{cc}

     			\includegraphics[scale=0.28]{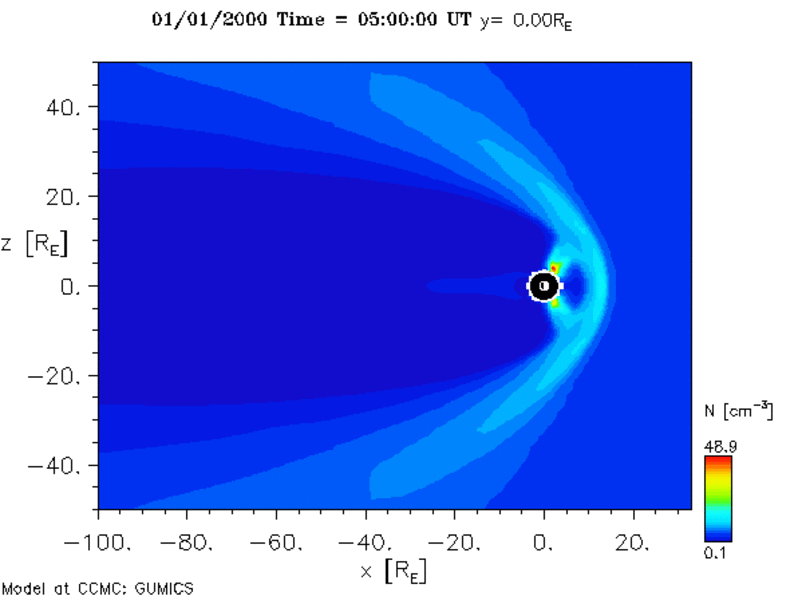}&
     			
     			\includegraphics[scale=0.24]{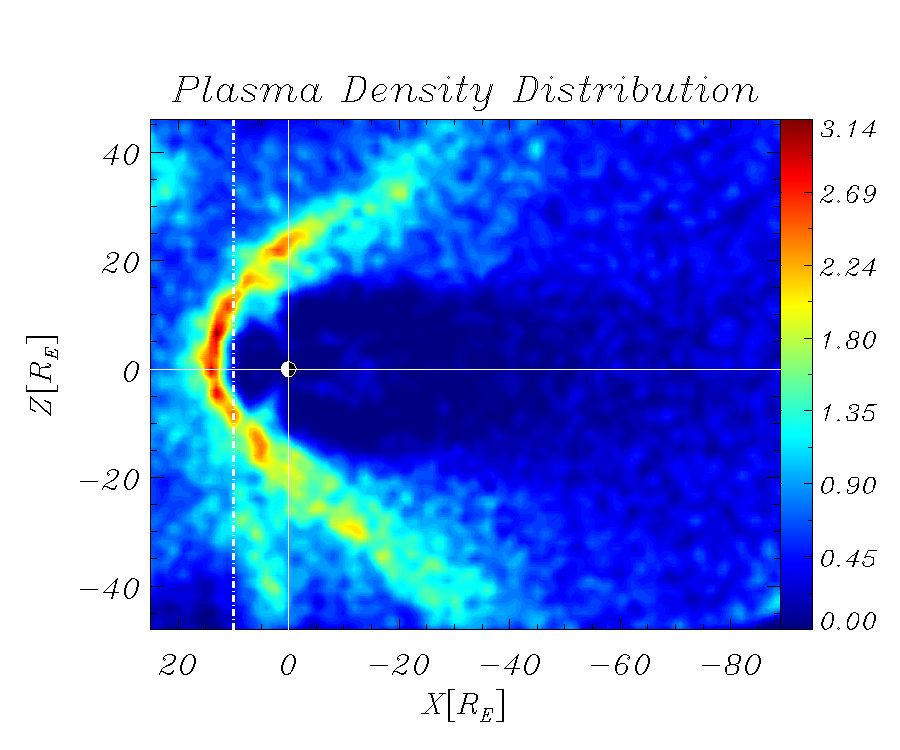}\\
     			
     			A&B \\

     		\end{tabular}
     		\caption{Plasma density distribution in Noon-Midnight axis. Panel \textrm{A} shows the \textrm{MHD} system generated plasma distribution, whilte panel \textrm{B} shows the plasma distribution as simulated by \textrm{PIC} code} \label{Fig:runs}
     		
     	\end{figure}

\begin{figure}[htp]
     		
     		\centering

     		\begin{tabular}{cc}
     			
     			
     			\includegraphics[width=72.5mm]{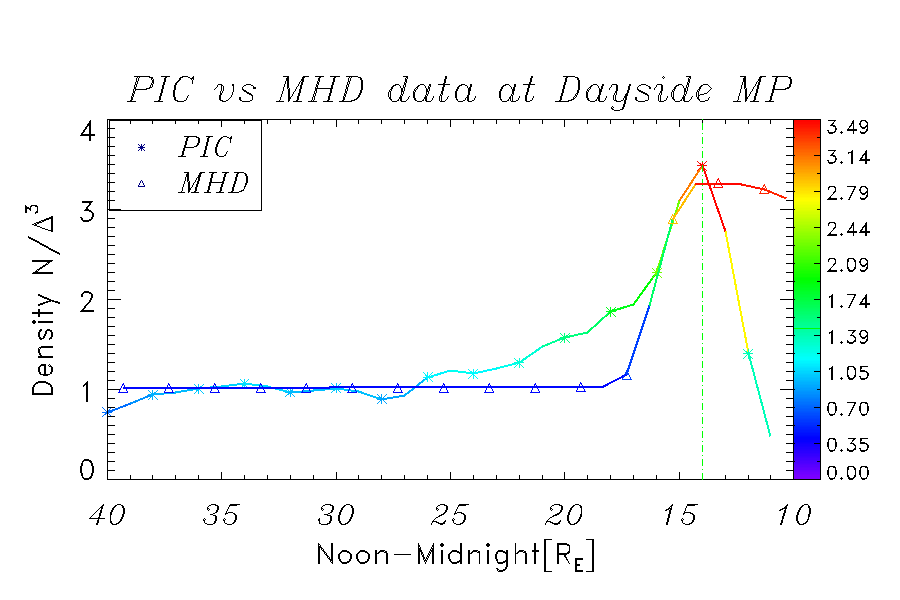}&
     			
     			\includegraphics[width=72.5mm]{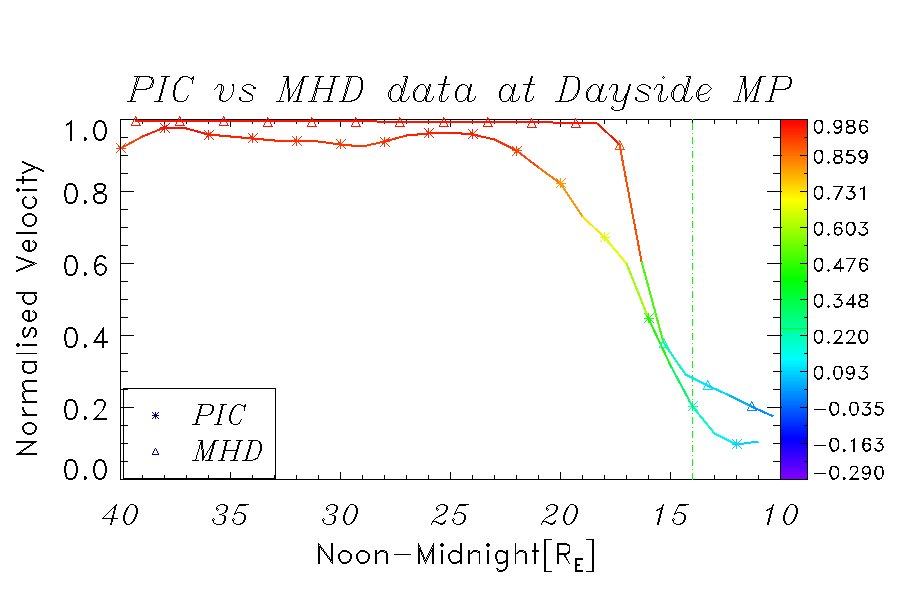}\\
     			A&B\\
     			\includegraphics[width=72.5mm]{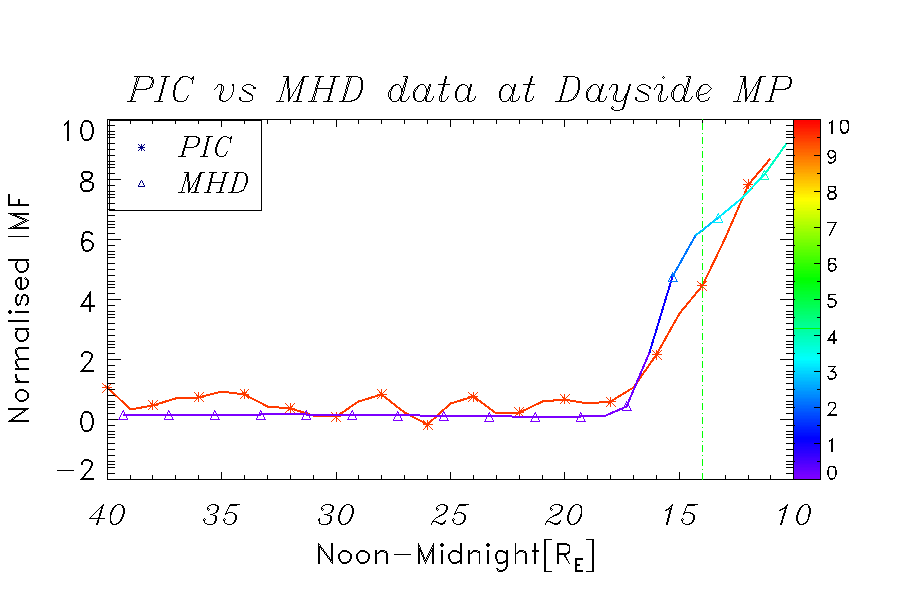}&
     			
     			\includegraphics[width=72.5mm]{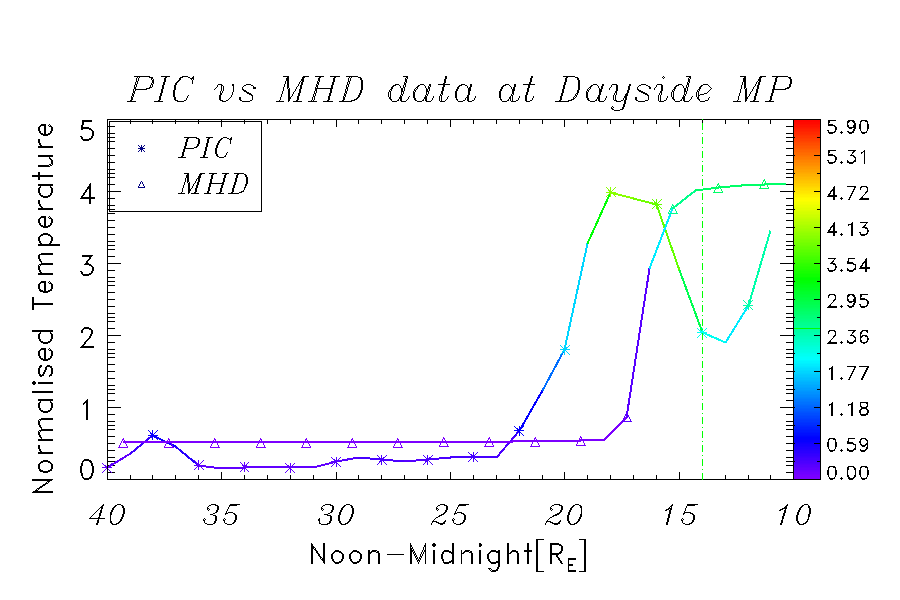}\\
     			C&D\\
     		\end{tabular}
     		\caption{ Colored Elevated plasma parameters plotted in nose direction. In frame (A) density profile  multi-plot is shown in the dayside magnetosphere as simulated by PIC code ($ \star $ symbol) and MHD ($ \triangle $symbol) . Similarly , Velocity, IMF and Temperature is depicted in Frames (B), (C) and (D) respectively. All parameters are normalized to their input values. Units are in $ R_{E} $.}\label{Fig:paramxz}
     		
     	\end{figure}

\begin{figure}[htp]
     		
     		\centering

     		\begin{tabular}{cc}
     			
     			
     			\includegraphics[width=72.5mm]{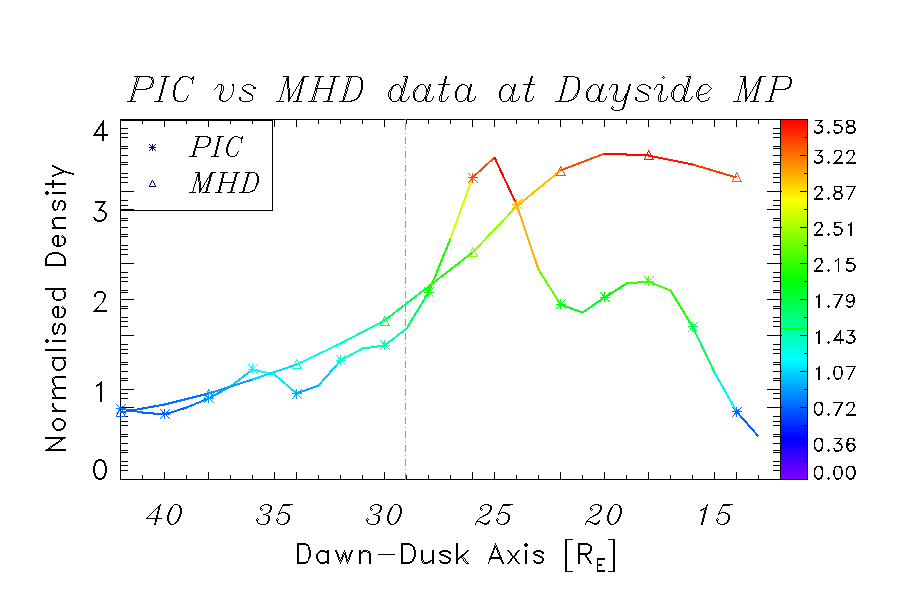}&
     			
     			\includegraphics[width=72.5mm]{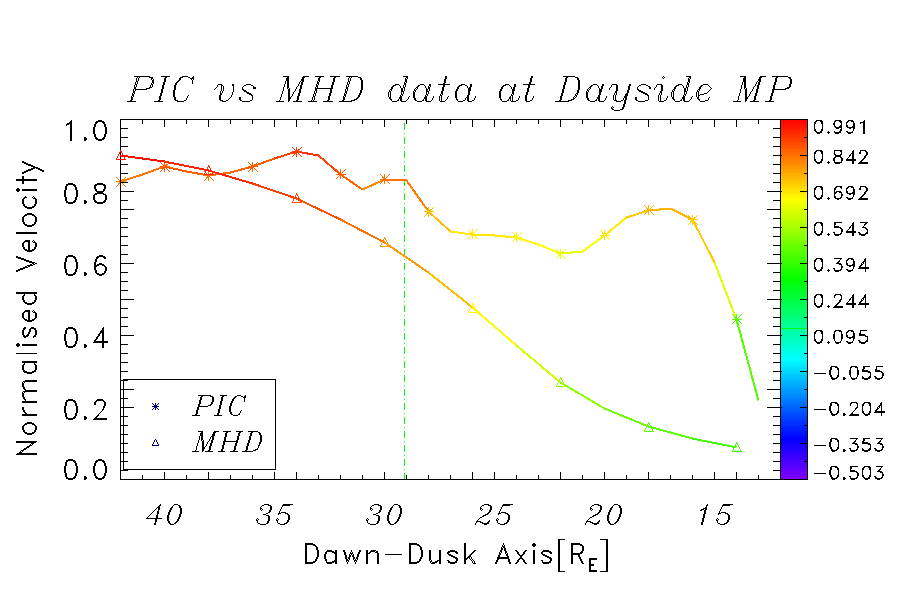}\\
     			A&B\\
     			\includegraphics[width=72.5mm]{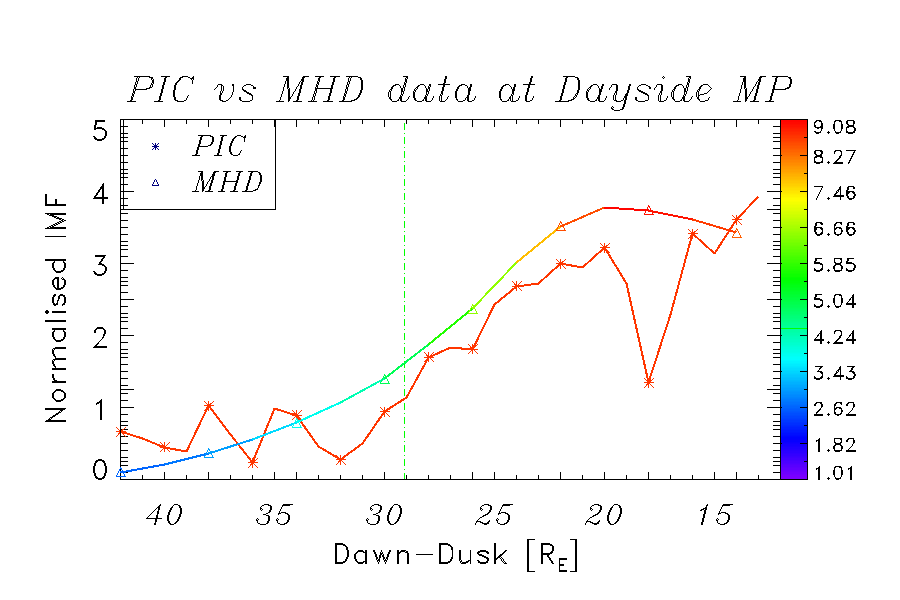}&
     			
     			\includegraphics[width=72.5mm]{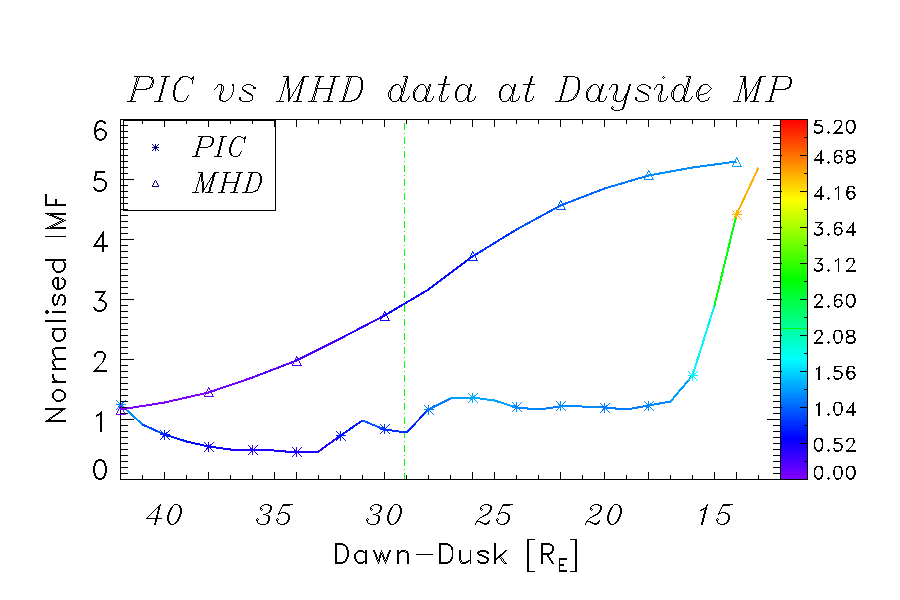}\\
     			C&D\\
     		\end{tabular}
     		\caption{ Colored Elevated plasma parameters plotted  in the dusk direction. In frame (A) density profile  multi-plot is shown in the dayside magnetosphere as simulated by PIC code ($ \star $ symbol) and MHD ($ \triangle $symbol) . Similarly , Velocity, IMF and Temperature is depicted in Frames (B), (C) and (D) respectively. All parameters are normalized to their input values. Units are in $ R_{E} $.} \label{Fig:paramxy}
     		
     	\end{figure}

     	The large scale Earth's magnetosphere   is simulated  by \textrm{PIC} EM relativistic code in parallel with \textrm{MHD}  code. One of the key features of both runs is  the structure, position and shape of the Earth's bow shock as depicted in the results. The geometry of the Earth's bow shock resembles bullet-like shape(see Fig. \ref{Fig:runs}). Its position was found by both codes to be equal to $ 14.9 R_{E} $ as measured along the nose direction from planet position\,, and $ 29. R_{E} $ along  the dusk direction.   These  results are in good agreement with  \textrm{in situ} measurements obtained for $ M_{A} $ values  within the range  ($ 8-13 R_{E} $) along ($ OX $ direction) as reported by \cite{Peredo1995}\,, and shown in Fig ( \ref{Fig:paramxz} panel A) and in Fig (\ref{Fig:paramxy} panel A).\\
     	On the other hand, we see in Fig \ref{Fig:paramxz} panel (B)) how the velocity simulation of both codes decreases and stagnated at the bow shock position. The  simulated velocity  by \textrm{PIC} code shows a spatial delay compared to the sharp decrease of  the \textrm{MHD} code\,, seemingly caused by the effect of thermal electrons in the foreshock region (i.e.\ electrons velocity spatial distribution in  Fig \ref{Fig:ionelec} panel A \& B). Same effect of the velocity profile of both codes can also be seen in Fig (\ref{Fig:paramxy} panel (B)) in the dusk side. The \textrm{IMF} profile along nose and dusk directions (Fig \ref{Fig:paramxz} and \ref{Fig:paramxy} panel C)  respectively, shows similarities between both codes in the behavior of the magnetic field at the bow shock position. \vskip0.2cm 
     	
     	On the other hand, both temperature profiles in Figs \ref{Fig:paramxz} and \ref{Fig:paramxy} panel (D) apparently show differences and spatial lags in the temperature jump. This is because in \textrm{PIC} code electrons temperature are included, but not in \textrm{MHD} codes, additionally the thermal velocities of electrons offsets their smaller masses. \vskip0.2cm 
     	
\begin{figure}[htp]
     		
     		\centering

     		\begin{tabular}{c}
     			
     			
     			\includegraphics[width=70mm]{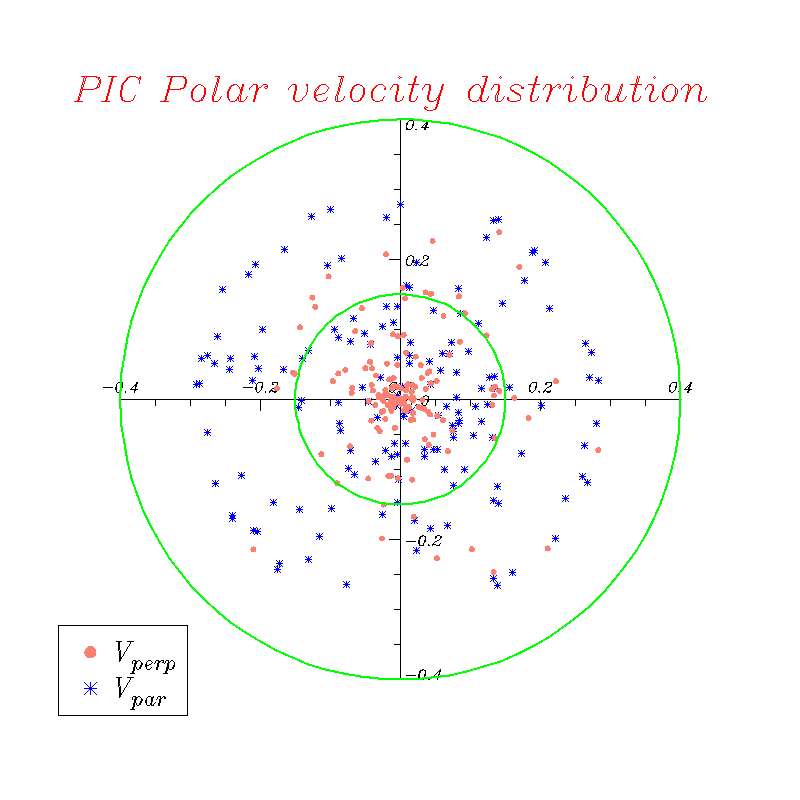}\\
     			\textbf{\large A}\\
     			\includegraphics[width=70mm]{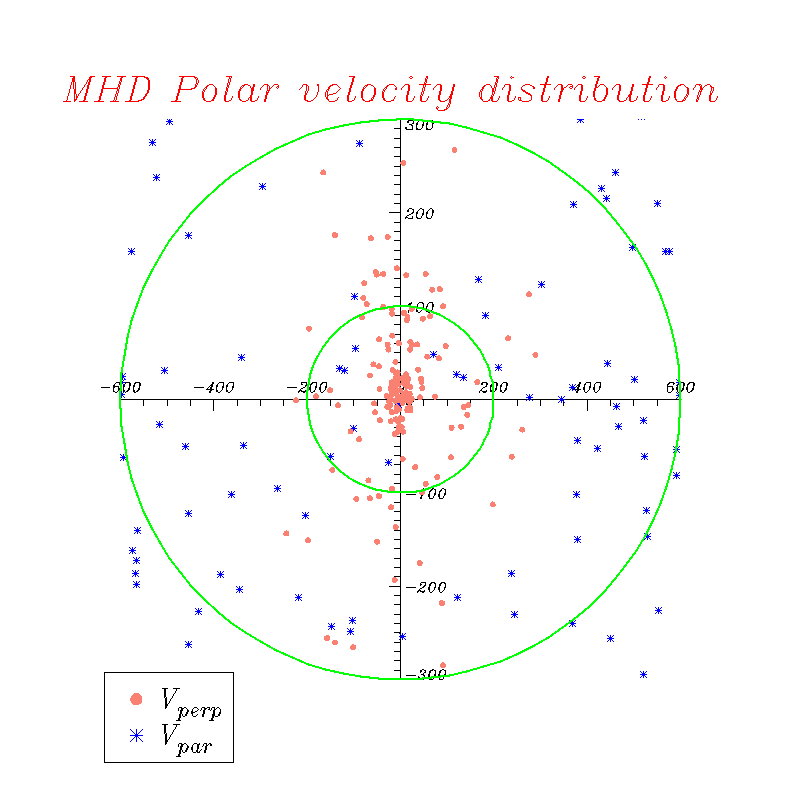}\\
     			
     			\textbf{\large B}\\

     		\end{tabular}
     		\caption{ Panel A shows the polar distribution of parallel and perpendicular velocities as generated by PIC code (input value is $ 0.25 \equiv 500 \quad km.sec^{-1}) $. Panel B shows same distribution as simulated by MHD code(input solar wind value is $500 \quad km.sec^{-1} $)} \label{Fig:polar}
     		
     	\end{figure}

     	To further show similarities and differences between the two codes, the parallel and perpendicular velocity distributions of the \textrm{PIC} code are polar-plotted in Fig \ref{Fig:polar} panel (A) and for \textrm{MHD} in panel (B). The maximum parallel velocity distribution values  is factor 3.4 than that of the perpendicular ones for the \textrm{PIC} code, whilst on the other hand this ratio is  factor 2 for the \textrm{MHD}.

\begin{figure}[htp]
     		
     		\centering

     		\begin{tabular}{c}
     			
     			
     			\includegraphics[width=120mm]{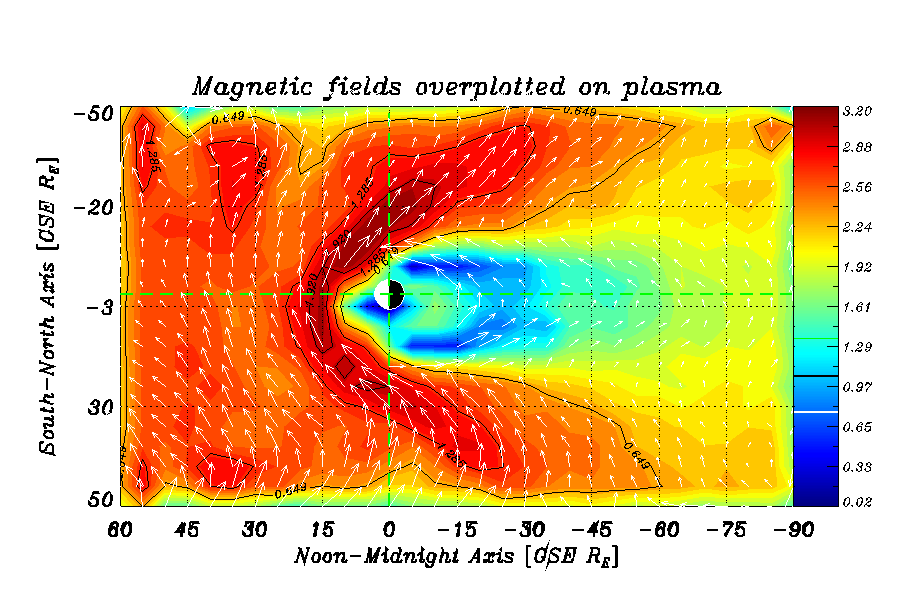}\\
     			\textbf{\large A}\\
     			\includegraphics[width=120mm]{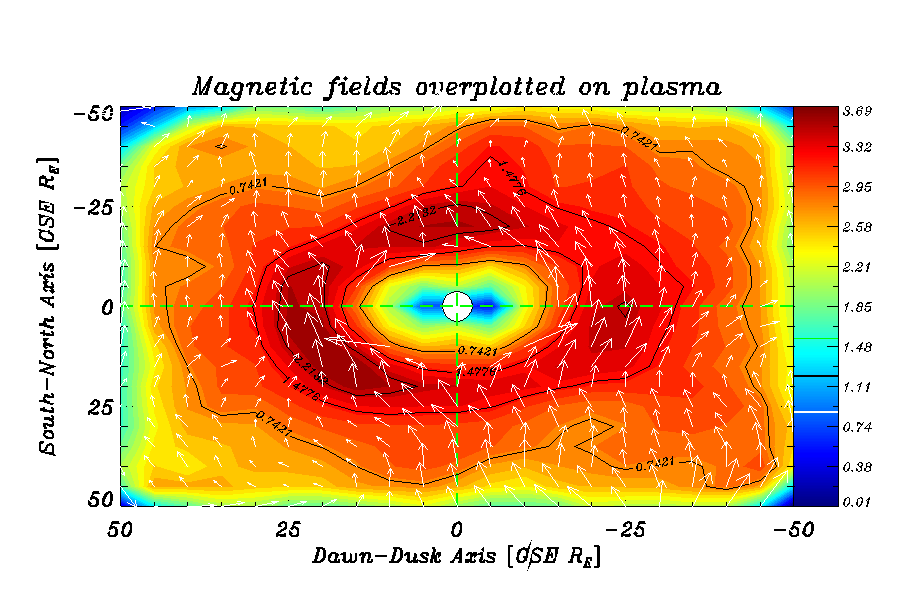}\\
     			
     			\textbf{\large B}\\

     		\end{tabular}
     		\caption{ Vector fields in $ X-Z $ direction in panel \textrm{A}  are over-plotted on \textrm{2D} density distribution. This  figure shows fields and particles updates after being run to step time $ 900 \Delta t $. In panel \textrm{B} vector fields are plotted in $Y-Z$ direction are over-plotted  on plasma. Note the magnetic field orientation at the foreshock region in both panels} \label{Fig:velovect}
     		
     	\end{figure}

     	On the other hand, If we base our diagnostic on the magnetosonic Mach number $ M_{MS} \approx 5 $ in both codes(see Table \ref{Tab:1}),  the above comparison shows that   \textrm{PIC} simulation can successfully recover the traditional results of the  \textrm{MHD} model in terms of the macrostructure of the bow shock, but at a much lower cost in computational time. Further more to our diagnostic, the magnetic field (northward input) is shown quasi perpendicularly oriented where it is plotted over the plasma density in Noon-Midnight in Fig  \ref{Fig:velovect} panel A, and in Dawn-Dusk direction in \ref{Fig:velovect} panel B.
     	\vskip 0.25cm

	\section{Analysis and Discussion}\label{secanal}
     	\vskip 0.25cm
     	Since the early models of the magnetosphere by \cite{Chapman1930b} through \cite{Dungey1962} until present, statistical, theoretical, observational, and modeling  have been extensively used  to comprehensively resolve the magnetospheric unsolved problems. 
     	
     	In our case, we don't re-invent the wheel. Our  code development has been considered  for upgrade for so many years and still in terms of spatial and temporal resolutions. Additional considerations are given to handle physical instabilities and to reduce CPU run time. In the near future, we will have a validated version of the code that is enhanced in terms of spatial and temporal resolutions with real ion to electron mass ratio. In order to keep the physical problem under investigation fixed, one has to adjust all other physical inputs parameters simultaneously. Because if one changes for example particle density to reduce statistical noises, then all other physical quantities will vary i.e.\  \cite{Cai2003}. This is exactly what has been taken care of in the current case study. 
     	The global structure of the collisionless bow shock was investigated by \cite{Omidi2005}, in their model ions are treated kinetically, whilst electrons are treated as a massless fluids. It is worth noting that they used 2.5D simulations. Two spatial dimensions and  3D for velocity and currents. Another work consider the magnetosphere simulation by 2.5D was reported by \cite{Moritaka2012}. They reproduce the magnetosphere. In a recent study by \cite{Cai2015}  a large scale 3D PIC code is used to study the whole terrestrial magnetosphere using ion to electron mass ratio equal to $\frac{1}{16}$.  In the current study a large scale structure of the magnetosphere was recovered but with full 3D simulations, in addition that electrons kinetics are included in the run. In their simulations and ours as well our physical units were scaled to ion inertial lengths and all were successful to recover the large scale magnetosphere.\\ \\
     	However the	\textrm{PIC} simulation is not a faithful representation of the plasma physics, it is still a must. On the other hand, even with the huge super-computing facilities available nowadays, it is quite impossible to simulate real magnetosphere. Thus scaling is an answer as quoted in the above references. After all these years and all these advances in the magnetospheric physics, we still don't know the magnetosphere(private communication recently with Mikhail Sitnov).  One can imagine a cuboid of volume of real magnetosphere  equal to $1.5 \times 10^6$  Earth Radii ($R_{E}^3$) is considered for simulation while one is looking for kinetic processes that take place in few 10s of meters. 
     	In this paper, a macro-structure of the Earth magnetosphere is successfully simulated. It is quite clear that we don't have a High-Definition(HD) image with the current scaled values and their corresponding spatial and temporal resolutions, but, for global structure a little blur image is enough to give a glimpse about the considered physical problem in hand.  I think if a comprehensive answer is reached in the space plasma physics field, it would have been enough for the community to pursue the discipline any further. We still on the long road to reach out there. 
     	In this section  we will analyze the criteria under which the \textrm{PIC} code is used in this study. The \textrm{MHD} code structure, boundary conditions  are well defined in  \citet{Janhunen2012}. Adopting the analysis  in \cite{Buechner2003},  we simulated a dynamic system  that include the bow shock in the macroscopic scale, we made sure that our total run time  is very much greater than the ion gyroperiod 
     	$\tau_{total} \gg \omega_{ci}^{-1}$, where $ \omega_{ci} $ is the upstream ion gyro-frequency. Typically the shock thickness is of order of few $R_{E}$, which is very much smaller than the plasma simulation box size.

\begin{figure}[htp]
     		
     		\centering

     		\begin{tabular}{c}
     			
     			
     			\includegraphics[width=120mm]{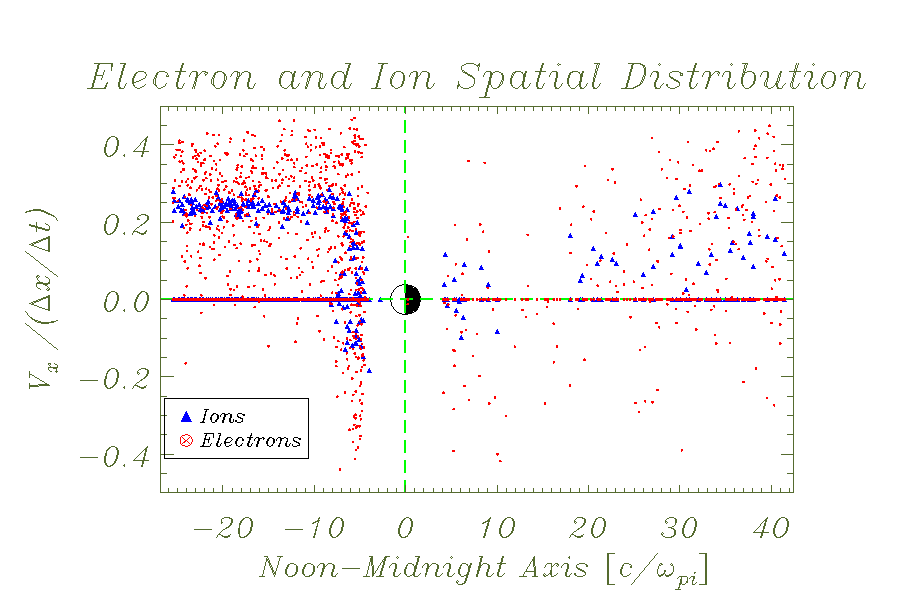}\\
     			\textbf{\large A}\\
     			\includegraphics[width=120mm]{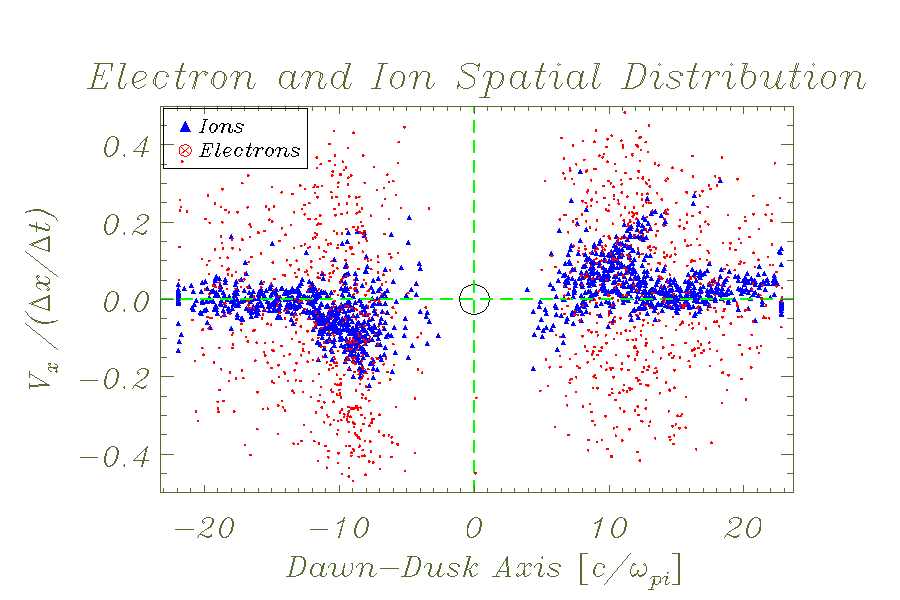}\\
     			
     			\textbf{\large B}\\

     		\end{tabular}
     		\caption{In panel \textrm{(A)} spatial distribution of ions and electrons velocities taken at nose direction both in day and night side of the magnetosphere. The thermal behavior of electrons can be clearly seen in this figure, especially at the day side portion of the magnetosphere. In panel \textrm{(B)}, the same spatial distribution but is taken in the dusk direction}\label{Fig:ionelec}
     		
     	\end{figure}

     	In Fig \ref{Fig:ionelec} panels \textrm{A} and \textrm{B} show the nose and the dusk direction respectively of  ions (in blue) and electrons (in red) velocity spatial distribution. In panel \textrm{A}, inflow of ions have relatively small velocity variations before it reached the shock terminal. Whilst on the other hand, the electrons velocity in the dayside spans high variations because of their thermal motions.  It is also worth noting the backstreaming of ions and electrons in the foreshock regions. The corresponding velocity spatial distributions in the dawn-dusk directions is shown in panel \textrm{B}. 
     	
     	On the other hand, we measured the velocity (thermal) of ions in the foreshock region  (at$  \sim 15 R_{E} $)  which is found to be equal to  0.10665 ($ \approx ~213.30 $ $ \mathrm{km.s^{-1}} $) and in the Magnetosheath  at around  $ 12 R_{E}  $  it  is  $ 0.03150  $    ($\approx   63.0 \mathrm{km.s^{-1} }$ ) , the reference value of solar wind speed is 0.25 ($ \approx ~500  \mathrm{km.s^{-1}} $). These findings are consistent with the recent study of  \cite{Pokhotelov2013}
     	
\begin{figure}[th]
     		\centering
     		\includegraphics[width=1.\linewidth]{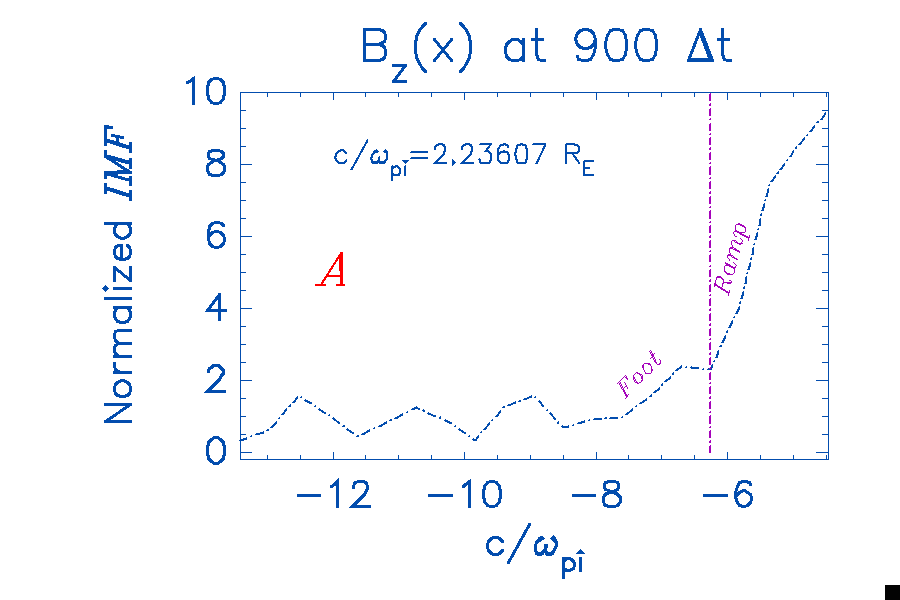}
     		\caption{shows the Bz(x) structure at the bow shock/foreshock region. Bz(x) as simulated by PIC code clearly depict the foot and ramp structure, compared with the theoretical model reported in \cite{Treumann2009} in Fig 10.}
     		\label{fig:treumannmodel}
     	\end{figure}

\begin{figure}[th]
     		\centering
     		\includegraphics[width=1.\linewidth]{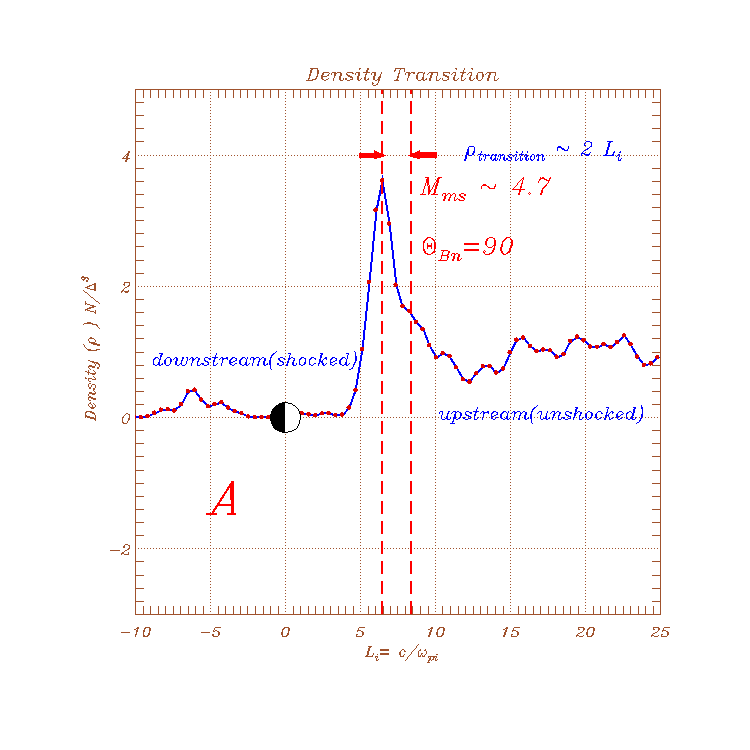}
     		\caption{shows the density transition between downstream (shocked) and upstream (unshocked) as simulated by our cod, the red vertical lines show the density transition scale. The figure is mirror imaged for comparison purposes. Our result is compared with cluster data density transition scale as reported by \cite{Bale2003} in Fig 5.4}
     		\label{fig:baleandmodel}
     	\end{figure}

     	Another result we report in the this paper is the 
     	magnetic field jump was zoomed in and plotted in the foreshock region Fig \ref{fig:treumannmodel}, where the foot and the ramp of the shock is shown. Overshoot of the shock didn't appear at this current version of the code. This result compared with analysis of the shock dynamics by \cite{Treumann2009}. Also another result we report here when the width of the density transition region of the shock was calculated and was found to be $\approx 2$ ion inertial lengths($ c/\omega_{pi} $) as in Fig \ref{fig:baleandmodel}. This result is in full agreement with \cite{Bale2003}.This figure is mirror-imaged for comparison reason.

     	The width of our ramp is 1.7 $ c/\omega_{pi} =L_{i}$\,, which is comparable to the value obtained in  \cite{Krasnoselskikh2013}  $\approx   1.4 L_{i}  $. 
     	
     	However, it is unambiguously established that many observed thinnest ramps are less than      $ 5 c/\omega_{pi} $ thick and there was  an apparent trend for lower values as $ \theta_{Bn} \longrightarrow 90^{0} $.
     	The plasma inertia effects is considered in our   \textrm{PIC} simulation, as a consequence the length of the simulation box is very much larger than the Debye length $\lambda _{Dei}$=$ (0.11,0.11) $, the gyro-radius $\rho _{ce,i}$=$(1.25,20)$ and the inertia lengths $\frac{c}{\omega_{pe,i}}= (5,80)$for electrons and ions respectively. 
     	\vskip 0.25cm
     	On the other hand, a quick look at Fig \ref{Fig:paramxz} ion density jump fact 3, and the foot of the shock clearly appears. A time sequence study of such shocks revealed by   \textrm{PIC}, should be carried out in a new paper for deeper verification of these preliminary  results, where in Fig \ref{Fig:paramxz} the shock is only shown at 900 ~$ \Delta t $.  
     	\vskip 0.25cm
     	
     	It is also worth noting that in Fig \ref{Fig:ionelec} the ions and electrons at the upstream of the bow shock\,, have high velocities, which is consistent with observation \cite{Filbert1979} and \cite{Fitzenreiter1984}.\\
     	One final point is that we can  follow the motion of
     	electrons and ions in the self-consistent ~$ \vec{E} $
     	and/or $ \vec{B} $~ fields obtained from a solution of Maxwell's equations, with relativistic
     	effects are readily included by the use of the Lorentz equation
     	of motion. At this level no approximations
     	in the basic laws of mechanics and electromagnetism is introduced, and thus the full range of collisionless plasma physics
     	is included in such a model \cite{Pritchett2000}\,, which is the case of the current study. 
     	\newpage

	\section{Conclusion}
     	The results of this study are summarized as the following:
     	\begin{enumerate}
     		\item The output data of both runs are  retrieved and normalized to input plasma parameters. In this paper, we show   distinct features: the bow shock  position , jump conditions,  plasma density, and fields distributions in specific geometric configurations.
     		\item Both codes have showed that the bow shock location is found to be at $ \sim 14 R_{E} $ along the Sun-Earth line and  at $\sim 29 R{_E}$ along the dawn-dusk direction, with a factor 3 in density jump. This result is consistent with in \textrm{situ} observations obtained during similar SW and IMF conditions. 
     		\item The  (thermal) velocity  of ions in the foreshock region  (at$  \sim 15 R_{E} $)  is  measured and found  to be  0.10665 ($ \approx ~213.30 $  $ \mathrm{km.s^{-1}}  $) and  its value in the  Magnetosheath  at around  $ 12 R_{E}  $  is  $ 0.03150  $    ($ \approx   63.0 \mathrm{km.s^{-1} }  $) , the reference value of solar wind speed is 0.25 ($ \approx ~500  \mathrm{km.s^{-1}} $).
     		\item The structure of the magnetic field jump at the shock  $ B_{z}(x)  $ of the foot and ramp of the magnetic field is obtained by the \textrm{PIC} code. The width of our ramp is 1.7 $ c/\omega_{pi} =L_{i}$ which is comparable to the value of  $\approx   1.4 L_{i}  $ \cite{Krasnoselskikh2013} . 
     		\item The density transition between the shocked plasma in the downstream and the  unshocked plasma in the upstream is found to be $ \approx 2 $  ion inertial length ($ c/\omega_{pi}  $)  at the magnetosonic number $ 4.7  $ when $ \Theta_{Bn}=90^{o} $
     		\item Both simulations reproduce the same basic macroscopic features of the Earth's magnetosphere. However, for the \textrm{PIC} code, a noisy current-sheet naturally appears, but it is absent in the   \textrm{MHD} results.
     		\item The velocity distribution of different species across and parallel  the ambient magnetic field can be derived anywhere in the magnetosphere from the \textrm{PIC} simulation, the derivation of that velocity distribution is also absent in   \textrm{MHD} results.
     		\item In \textrm{PIC} models one can follow the motion of electrons and ions in the self-consistent ~$ \mathrm{\vec{E} }$ and/or $ \mathrm{\vec{B}} $~ fields obtained from the solution of Maxwell's equations, with relativistic effects are readily included by the use of the Lorentz equation of motion. 
     		\item In contrast, macroscopic properties of the magnetosphere obtained from   \textrm{MHD} simulations can be directly compared to observations, while only scaled quantities from   \textrm{PIC} simulations are useful in such comparisons.
     		\item The results obtained thus far from the present study strongly suggest using   \textrm{MHD} and \textrm{PIC} codes in a complementary manner as a new strategy for better understanding of the magnetosphere-solar wind system.
     		\item The \textrm{PIC} showed the backstreaming velocity distribution of both ions and electrons on the nose and  on the dusk-direction in the dayside magnetosphere (foreshock, transition shock, magnetosheath and in the magnetotail).
     		\item This working version of our \textrm{PIC} code is powerful to simulate large scale magnetospheric electrodynamics. It is undoubtedly capable of simulating more sophisticated kinetics, such as reconnection, cusp dynamics and current systems if and only if a better computer resources and multiprocessors super computing facilities are available, in order to be able to reduce grid cell size and to increase the number of pair particles of the simulation box. 
     		
     	\end{enumerate}
     	%%%%%%%%%%%
     	%%%%%%%%%%%
     	
	\section{acknowledgements}
     	This work would have not been done without the insights and hard work on code development  by Dr Lotfi Ben-Jaffel,IAP. I would like also to thank the IAP-CNRS, Paris, France. The author also thank to David Sibeck of NGFC-NASA,  Bob Clauer of VT, and Douglas Staley President of NIA for their continuous supports and insights. The Author also would like to thank Zamala program and Bank of Palestine for supporting my research visits to the US.

     	%%%%%%%%%%%
     	%%%%%%%%%%%
     	%%%%%%%%%%%%%
     	%%%%%%%%%%%%
     
     	%%%%%%%%%%%%%

     	\clearpage
     	\singlespacing
     	\bibliographystyle{abbrvnat}
     	\bibliography{../baraka}

\begin{thebibliography}{66}
\providecommand{\natexlab}[1]{#1}
\providecommand{\url}[1]{\texttt{#1}}
\expandafter\ifx\csname urlstyle\endcsname\relax
  \providecommand{\doi}[1]{doi: #1}\else
  \providecommand{\doi}{doi: \begingroup \urlstyle{rm}\Url}\fi

\bibitem[Axford(1962)]{Axford1962}
W.~Axford.
\newblock The interaction between the solar wind and the earth's magnetosphere.
\newblock \emph{J. Geophys. Res.}, 67\penalty0 (10):\penalty0 3791--3796, 1962.

\bibitem[Bale et~al.(2003)Bale, Mozer, and Horbury]{Bale2003}
S.~Bale, F.~Mozer, and T.~Horbury.
\newblock Density-{Transition} {Scale} at {Quasiperpendicular} {Collisionless}
  {Shocks}.
\newblock \emph{Phys. Rev. Lett.}, 91\penalty0 (26):\penalty0 265004, Dec.
  2003.
\newblock ISSN 0031-9007.
\newblock \doi{10.1103/PhysRevLett.91.265004}.
\newblock URL \url{http://link.aps.org/doi/10.1103/PhysRevLett.91.265004}.

\bibitem[Baraka(2007)]{Baraka2007}
S.~Baraka.
\newblock \emph{Etude de l'interactionentre le vent solaire et la magnetosphere
  de la {Terre}: {Modele} theorique et {Application} sur l'analyse de donnees
  de l'evenement du {Halloween} d'octobre 2003}.
\newblock PhD thesis, Université Pierre et Marie Curie-Paris VI, 2007.
\newblock URL \url{https://tel.archives-ouvertes.fr/tel-00138416/en/}.

\bibitem[Baraka and Ben-Jaffel(2007)]{Baraka2007b}
S.~Baraka and L.~Ben-Jaffel.
\newblock Sensitivity of the {Earth}'s magnetosphere to solar wind activity:
  {Three}-dimensional macroparticle model.
\newblock \emph{Journal of Geophysical Research (Space Physics)}, 112:\penalty0
  6212, June 2007.
\newblock \doi{10.1029/2006JA011946}.

\bibitem[Baraka and Ben-Jaffel(2011)]{Baraka2011}
S.~Baraka and L.~Ben-Jaffel.
\newblock Impact of solar wind depression on the dayside magnetosphere under
  northward interplanetary magnetic field.
\newblock \emph{Annales Geophysicae}, 29:\penalty0 31--46, Jan. 2011.
\newblock \doi{10.5194/angeo-29-31-2011}.

\bibitem[Baraka and Jaffel(2014)]{Baraka2014}
S.~Baraka and L.~Jaffel.
\newblock 3d {Simulation} of {Eareth}'s magentosphere by {Particle}-{In}-{Cell}
  and magnetohydrodynamics models: parameteric study.
\newblock In \emph{{AGU} {Fall} {Meeting} {Abstracts}}, volume~1, page 4222,
  2014.

\bibitem[Baraka et~al.(2013)Baraka, Jaffel, and Dandouras]{Baraka2013}
S.~Baraka, L.~Jaffel, and I.~Dandouras.
\newblock The unusual event of {Jan} 21st 2005 observed by {Cluster}
  spacecracts is considered for comparison by {PIC} {EM} {Relativistic} code
  simulation.
\newblock In \emph{{AGU} {Fall} {Meeting} {Abstracts}}, volume~1, page 2236,
  2013.

\bibitem[Ben-Jaffel and Ballester(2014)]{Ben-Jaffel2014}
L.~Ben-Jaffel and G.~E. Ballester.
\newblock Transit of exomoon plasma tori: new diagnosis.
\newblock \emph{The Astrophysical Journal Letters}, 785\penalty0 (2):\penalty0
  L30, 2014.

\bibitem[Ben-Jaffel et~al.(2013)Ben-Jaffel, Strumik, Ratkiewicz, and
  Grygorczuk]{Ben-Jaffel2013d}
L.~Ben-Jaffel, M.~Strumik, R.~Ratkiewicz, and J.~Grygorczuk.
\newblock The {Existence} and {Nature} of the {Interstellar} {Bow} {Shock}.
\newblock \emph{{\textbackslash}apj}, 779:\penalty0 130, Dec. 2013.
\newblock \doi{10.1088/0004-637X/779/2/130}.

\bibitem[Birdsall and Langdon(2005)]{Birdsall2005}
C.~K. Birdsall and A.~B. Langdon.
\newblock \emph{Plasma {Physics} {Via} {Computer} {Simulaition}}.
\newblock CRC Press, 2005.

\bibitem[Bonifazi and Moreno(1981)]{Bonifazi1981}
C.~Bonifazi and G.~Moreno.
\newblock Reflected and diffuse ions backstreaming from the {Earth}'s bow shock
  2. {Origin}.
\newblock \emph{Journal of Geophysical Research: Space Physics (1978–2012)},
  86\penalty0 (A6):\penalty0 4405--4413, 1981.

\bibitem[Buneman(1993)]{Buneman1993}
O.~Buneman.
\newblock Computer space plasma physics.
\newblock \emph{Simulation Techniques and Software}, page~67, 1993.

\bibitem[Buneman et~al.(1980)Buneman, Barnes, Green, and Nielsen]{Buneman1980}
O.~Buneman, C.~Barnes, J.~Green, and D.~Nielsen.
\newblock Principles and capabilities of 3-{D}, {EM} particle simulations.
\newblock \emph{J. Comput. Phys.}, 38\penalty0 (1):\penalty0 1--44, 1980.

\bibitem[Buneman et~al.(1992)Buneman, Neubert, and Nishikawa]{Buneman1992}
O.~Buneman, T.~Neubert, and K.-I. Nishikawa.
\newblock Solar wind-magnetosphere interaction as simulated by a 3-{D} {EM}
  particle code.
\newblock \emph{Plasma Science, IEEE Transactions on}, 20\penalty0
  (6):\penalty0 810--816, 1992.

\bibitem[Buneman et~al.(1995)Buneman, Nishikawa, and Neubert]{Buneman1995}
O.~Buneman, K.-I. Nishikawa, and T.~Neubert.
\newblock Solar {Wind}-{Magnetosphere} {Interaction} as {Simulated} by a 3d,
  {Em} {Particle} {Code}.
\newblock \emph{Space Plasmas: Coupling Between Small and Medium Scale
  Processes}, pages 347--356, 1995.

\bibitem[Bykov and Treumann(2011)]{Bykov2011}
A.~Bykov and R.~Treumann.
\newblock Fundamentals of collisionless shocks for astrophysical application,
  2. relativistic shocks.
\newblock \emph{The Astronomy and Astrophysics Review}, 19\penalty0
  (1):\penalty0 1--67, 2011.

\bibitem[Büchner et~al.(2003)Büchner, Dum, and Scholer]{Buechner2003}
J.~Büchner, C.~Dum, and M.~Scholer.
\newblock \emph{Space plasma simulation}, volume 615.
\newblock Springer, 2003.

\bibitem[Cai et~al.(2003)Cai, Li, Nishikawa, Xiao, Yan, and Pu]{Cai2003}
D.~Cai, Y.~Li, K.-I. Nishikawa, C.~Xiao, X.~Yan, and Z.~Pu.
\newblock Parallel 3-d electromagnetic particle code using high performance
  fortran: Parallel tristan.
\newblock In \emph{Space Plasma Simulation}, pages 25--53. Springer, 2003.

\bibitem[Cai et~al.(2006)Cai, Yan, Nishikawa, and Lembège]{Cai2006}
D.~Cai, X.~Yan, K.-I. Nishikawa, and B.~Lembège.
\newblock Particle entry into the inner magnetosphere during duskward {IMF}
  {By}: {Global} three-dimensional electromagnetic full particle simulations.
\newblock \emph{Geophys. Res. Lett.}, 33\penalty0 (12), 2006.

\bibitem[Cai et~al.(2015)Cai, Esmaeili, Lemb{\`e}ge, and Nishikawa]{Cai2015}
D.~Cai, A.~Esmaeili, B.~Lemb{\`e}ge, and K.-I. Nishikawa.
\newblock Cusp dynamics under northward imf using three-dimensional global
  particle-in-cell simulations.
\newblock \emph{Journal of Geophysical Research: Space Physics}, 120\penalty0
  (10):\penalty0 8368--8386, 2015.

\bibitem[Chapman and Ferraro(1930)]{Chapman1930b}
S.~Chapman and V.~C.~A. Ferraro.
\newblock A {New} {Theory} of {Magnetic} {Storms}.
\newblock \emph{{\textbackslash}nat}, 126:\penalty0 129--130, July 1930.
\newblock \doi{10.1038/126129a0}.

\bibitem[Dmitriev et~al.(2003)Dmitriev, Chao, and Wu]{Dmitriev2003}
A.~Dmitriev, J.~Chao, and D.~Wu.
\newblock Comparative study of bow shock models using {Wind} and {Geotail}
  observations.
\newblock \emph{J. Geophys. Res.}, 108\penalty0 (A12):\penalty0 1464, 2003.

\bibitem[Dungey(1962)]{Dungey1962}
J.~Dungey.
\newblock The interplanetary field and auroral theory.
\newblock \emph{Journal of the Physical Society of Japan Supplement},
  17:\penalty0 15, 1962.

\bibitem[Ellison et~al.(1993)Ellison, Giacalone, Burgess, and
  Schwartz]{Ellison1993}
D.~C. Ellison, J.~Giacalone, D.~Burgess, and S.~Schwartz.
\newblock Simulations of particle acceleration in parallel shocks: {Direct}
  comparison between {Monte} {Carlo} and one-dimensional hybrid codes.
\newblock \emph{Journal of Geophysical Research: Space Physics (1978–2012)},
  98\penalty0 (A12):\penalty0 21085--21093, 1993.

\bibitem[Filbert and Kellogg(1979)]{Filbert1979}
P.~C. Filbert and P.~J. Kellogg.
\newblock Electrostatic noise at the plasma frequency beyond the earth's bow
  shock.
\newblock \emph{{\textbackslash}jgr}, 84:\penalty0 1369--1381, Apr. 1979.
\newblock \doi{10.1029/JA084iA04p01369}.

\bibitem[Fitzenreiter et~al.(1984)Fitzenreiter, Klimas, and
  Scudder]{Fitzenreiter1984}
R.~J. Fitzenreiter, A.~J. Klimas, and J.~D. Scudder.
\newblock Detection of bump-on-tail reduced electron velocity distributions at
  the electron foreshock boundary.
\newblock \emph{{\textbackslash}grl}, 11:\penalty0 496--499, May 1984.
\newblock \doi{10.1029/GL011i005p00496}.

\bibitem[Fontaine et~al.(2015)Fontaine, Turc, and Savoini]{Fontaine2015}
D.~Fontaine, L.~Turc, and P.~Savoini.
\newblock Multipoint observations and simulation of the effects of {Earth}'s
  bow shock on magnetic clouds' structure and geoeffectiveness.
\newblock In \emph{{EGU} {General} {Assembly} {Conference} {Abstracts}},
  volume~17, page 5908, 2015.
\newblock URL \url{http://adsabs.harvard.edu/abs/2015EGUGA..17.5908F}.

\bibitem[Gombosi et~al.(2000)Gombosi, Zeeuw, and Groth]{Gombosi2000}
T.~Gombosi, D.~D. Zeeuw, and C.~Groth.
\newblock Global {MHD} modeling of space weather.
\newblock \emph{IEEE Trans, Plasma Sci., {\textbackslash}ldots}, 2000.
\newblock URL
  \url{http://scholar.google.com/scholar?hl=en&q=T.+I.+Gombosi%2C+D.+L.+De+Zeeuw%2C+C.+P.+T.+Groth%2C+K.+G.+Powell%2C+A.+J.+Ridley%2C+and+P.+Song%2C+%E2%80%9CGlobal+MHD+modeling+of+space+weather%2C%E2%80%9D+IEEE+Trans.+Plasma+Sci.%2C+2000&btnG=&as_sdt=1%2C5&as_sdtp=#0}.

\bibitem[Janhunen et~al.(2012)Janhunen, Palmroth, Laitinen, Honkonen, Juusola,
  Facsko, and Pulkkinen]{Janhunen2012}
P.~Janhunen, M.~Palmroth, T.~Laitinen, I.~Honkonen, L.~Juusola, G.~Facsko, and
  T.~Pulkkinen.
\newblock The {GUMICS}-4 global {MHD} magnetosphere–ionosphere coupling
  simulation.
\newblock \emph{Journal of Atmospheric and Solar-Terrestrial Physics},
  80:\penalty0 48--59, 2012.

\bibitem[Jelínek et~al.(2010)Jelínek, Němeček, Šafránková, Shue,
  Suvorova, and Sibeck]{Jelinek2010}
K.~Jelínek, Z.~Němeček, J.~Šafránková, J.-H. Shue, A.~V. Suvorova, and
  D.~G. Sibeck.
\newblock Thin magnetosheath as a consequence of the magnetopause deformation:
  {THEMIS} observations.
\newblock \emph{Journal of Geophysical Research (Space Physics)}, 115:\penalty0
  10203, 2010.
\newblock ISSN 0148-0227.
\newblock URL \url{http://adsabs.harvard.edu/abs/2010JGRA..11510203J}.

\bibitem[Keika et~al.(2009)Keika, Nakamura, Baumjohann, Angelopoulos, Kabin,
  Glassmeier, Sibeck, Magnes, Auster, Fornaçon, McFadden, Carlson, Lucek,
  Carr, Dandouras, and Rankin]{Keika2009}
K.~Keika, R.~Nakamura, W.~Baumjohann, V.~Angelopoulos, K.~Kabin, K.~H.
  Glassmeier, D.~G. Sibeck, W.~Magnes, H.~U. Auster, K.~H. Fornaçon, J.~P.
  McFadden, C.~W. Carlson, E.~A. Lucek, C.~M. Carr, I.~Dandouras, and
  R.~Rankin.
\newblock Deformation and evolution of solar wind discontinuities through their
  interactions with the {Earth}'s bow shock.
\newblock \emph{Journal of Geophysical Research (Space Physics)}, 114, 2009.
\newblock ISSN 0148-0227.
\newblock URL \url{http://adsabs.harvard.edu/abs/2009JGRA..114.0C26K}.

\bibitem[Kellogg(1962)]{Kellogg1962}
P.~J. Kellogg.
\newblock Flow of plasma around the earth.
\newblock \emph{Journal of Geophysical Research}, 67\penalty0 (10):\penalty0
  3805--3811, 1962.

\bibitem[Kowal et~al.(2011)Kowal, Dal~Pino, and Lazarian]{Kowal2011}
G.~Kowal, E.~d.~G. Dal~Pino, and A.~Lazarian.
\newblock Magnetohydrodynamic simulations of reconnection and particle
  acceleration: three-dimensional effects.
\newblock \emph{The Astrophysical Journal}, 735\penalty0 (2):\penalty0 102,
  2011.

\bibitem[Krasnoselskikh et~al.(2013)Krasnoselskikh, Balikhin, Walker, Schwartz,
  Sundkvist, Lobzin, Gedalin, Bale, Mozer, Soucek, Hobara, and
  Comisel]{Krasnoselskikh2013}
V.~Krasnoselskikh, M.~Balikhin, S.~N. Walker, S.~Schwartz, D.~Sundkvist,
  V.~Lobzin, M.~Gedalin, S.~D. Bale, F.~Mozer, J.~Soucek, Y.~Hobara, and
  H.~Comisel.
\newblock The {Dynamic} {Quasiperpendicular} {Shock}: {Cluster} {Discoveries}.
\newblock \emph{Space Sci. Rev.}, 178\penalty0 (2-4):\penalty0 535--598, Mar.
  2013.
\newblock ISSN 0038-6308.
\newblock \doi{10.1007/s11214-013-9972-y}.
\newblock URL \url{http://link.springer.com/10.1007/s11214-013-9972-y}.

\bibitem[Kronberg et~al.(2011)Kronberg, Bučík, Haaland, Klecker, Keika,
  Desai, Daly, Yamauchi, Gómez-Herrero, and Lui]{Kronberg2011}
E.~Kronberg, R.~Bučík, S.~Haaland, B.~Klecker, K.~Keika, M.~Desai, P.~Daly,
  M.~Yamauchi, R.~Gómez-Herrero, and A.~Lui.
\newblock On the origin of the energetic ion events measured upstream of the
  {Earth}'s bow shock by {STEREO}, {Cluster}, and {Geotail}.
\newblock \emph{Journal of Geophysical Research: Space Physics (1978–2012)},
  116\penalty0 (A2), 2011.

\bibitem[Kullen et~al.(2004)Kullen, Janhunen, and {others}]{Kullen2004}
A.~Kullen, P.~Janhunen, and {others}.
\newblock Relation of polar auroral arcs to magnetotail twisting and {IMF}
  rotation: {A} systematic {MHD} simulation study.
\newblock In \emph{Annales {Geophysicae}}, volume~22, pages 951--970, 2004.

\bibitem[Leboeuf et~al.(1978)Leboeuf, Tajima, Kennel, and Dawson]{Leboeuf1978a}
J.~Leboeuf, T.~Tajima, C.~F. Kennel, and J.~Dawson.
\newblock Global simulation of the time-dependent magnetosphere.
\newblock \emph{Geophys. Res. Lett.}, 5\penalty0 (7):\penalty0 609--612, 1978.

\bibitem[Lindman(1975)]{Lindman1975}
E.~Lindman.
\newblock “{Free}-space” boundary conditions for the time dependent wave
  equation.
\newblock \emph{J. Comput. Phys.}, 18\penalty0 (1):\penalty0 66--78, 1975.

\bibitem[Mann et~al.(2008)Mann, Milling, Rae, Ozeke, Kale, Kale, Murphy,
  Parent, Usanova, Pahud, et~al.]{Mann2008}
I.~Mann, D.~Milling, I.~Rae, L.~Ozeke, A.~Kale, Z.~Kale, K.~Murphy, A.~Parent,
  M.~Usanova, D.~Pahud, et~al.
\newblock The upgraded carisma magnetometer array in the themis era.
\newblock \emph{Space Science Reviews}, 141\penalty0 (1-4):\penalty0 413--451,
  2008.

\bibitem[Maynard et~al.(2011)Maynard, Farrugia, Burke, Ober, Scudder, Mozer,
  Russell, Rème, Mouikis, and Siebert]{Maynard2011}
N.~C. Maynard, C.~J. Farrugia, W.~J. Burke, D.~M. Ober, J.~D. Scudder, F.~S.
  Mozer, C.~T. Russell, H.~Rème, C.~Mouikis, and K.~D. Siebert.
\newblock Interactions of the heliospheric current and plasma sheets with the
  bow shock: {Cluster} and {Polar} observations in the magnetosheath.
\newblock \emph{Journal of Geophysical Research: Space Physics (1978–2012)},
  116\penalty0 (A1), 2011.

\bibitem[Meziane et~al.(2007)Meziane, Wilber, Hamza, Mazelle, Parks, Reme, and
  Lucek]{Meziane2007}
K.~Meziane, M.~Wilber, A.~Hamza, C.~Mazelle, G.~Parks, H.~Reme, and E.~Lucek.
\newblock Evidence for a high-energy tail associated with foreshock
  field-aligned beams.
\newblock \emph{J. Geophys. Res.}, 112\penalty0 (A1):\penalty0 A01101, 2007.

\bibitem[Meziane et~al.(2015)Meziane, Hamza, Maksimovic, and
  Alrefay]{Meziane2015}
K.~Meziane, A.~M. Hamza, M.~Maksimovic, and T.~Y. Alrefay.
\newblock The {Earth}'s bow shock velocity distribution function.
\newblock \emph{Journal of Geophysical Research: Space Physics}, 120\penalty0
  (2):\penalty0 1229--1237, 2015.
\newblock URL
  \url{http://onlinelibrary.wiley.com/doi/10.1002/2014JA020772/pdf}.

\bibitem[Moritaka et~al.(2012)Moritaka, Kajimura, Usui, Matsumoto, Matsui, and
  Shinohara]{Moritaka2012}
T.~Moritaka, Y.~Kajimura, H.~Usui, M.~Matsumoto, T.~Matsui, and I.~Shinohara.
\newblock Momentum transfer of solar wind plasma in a kinetic scale
  magnetosphere.
\newblock \emph{Physics of Plasmas (1994-present)}, 19\penalty0 (3):\penalty0
  032111, 2012.

\bibitem[Nishikawa(1997)]{Nishikawa1997}
K.-I. Nishikawa.
\newblock Particle entry into the magnetosphere with a southward interplanetary
  magnetic field studied by a three-dimensional electromagnetic particle code.
\newblock \emph{Journal of Geophysical Research: Space Physics (1978–2012)},
  102\penalty0 (A8):\penalty0 17631--17641, 1997.

\bibitem[Omidi et~al.(2005)Omidi, Blanco-Cano, and Russell]{Omidi2005}
N.~Omidi, X.~Blanco-Cano, and C.~Russell.
\newblock Macrostructure of collisionless bow shocks: 1. {Scale} lengths.
\newblock \emph{Journal of Geophysical Research: Space Physics (1978–2012)},
  110\penalty0 (A12), 2005.

\bibitem[Omidi et~al.(2013)Omidi, Sibeck, Blanco-Cano, Rojas-Castillo, Turner,
  Zhang, and Kajdič]{Omidi2013}
N.~Omidi, D.~Sibeck, X.~Blanco-Cano, D.~Rojas-Castillo, D.~Turner, H.~Zhang,
  and P.~Kajdič.
\newblock Dynamics of the foreshock compressional boundary and its connection
  to foreshock cavities.
\newblock \emph{Journal of Geophysical Research (Space Physics)}, 118:\penalty0
  823--831, 2013.
\newblock ISSN 0148-0227.
\newblock URL \url{http://adsabs.harvard.edu/abs/2013JGRA..118..823O}.

\bibitem[Palmroth et~al.(2002)Palmroth, Pulkkinen, Janhunen, and
  Wu]{Palmroth2002}
M.~Palmroth, T.~Pulkkinen, P.~Janhunen, and C.-C. Wu.
\newblock Stormtime energy transfer in global {MHD} simulation.
\newblock \emph{J. Geophys. Res.}, 108\penalty0 (A1):\penalty0 SMP24--1, 2002.

\bibitem[Palmroth et~al.(2005)Palmroth, Janhunen, Pulkkinen, Aksnes, Lu,
  Østgaard, Watermann, Reeves, and Germany]{Palmroth2005}
M.~Palmroth, P.~Janhunen, T.~Pulkkinen, A.~Aksnes, G.~Lu, N.~Østgaard,
  J.~Watermann, G.~Reeves, and G.~Germany.
\newblock Assessment of ionospheric {Joule} heating by {GUMICS}-4 {MHD}
  simulation, {AMIE}, and satellite-based statistics: towards a synthesis.
\newblock In \emph{Annales {Geophysicae}}, volume~23, pages 2051--2068.
  Copernicus GmbH, 2005.

\bibitem[Parks(2004)]{Parks2004}
G.~K. Parks.
\newblock Why space physics needs to go beyond the {MHD} box.
\newblock \emph{Space science reviews}, 113\penalty0 (1-2):\penalty0 97--121,
  2004.

\bibitem[Paschmann et~al.(1981)Paschmann, Sckopke, Papamastorakis, Asbridge,
  Bame, and Gosling]{Paschmann1981}
G.~Paschmann, N.~Sckopke, I.~Papamastorakis, J.~Asbridge, S.~Bame, and
  J.~Gosling.
\newblock Characteristics of reflected and diffuse ions upstream from the
  earth's bow shock.
\newblock \emph{Journal of Geophysical Research: Space Physics (1978–2012)},
  86\penalty0 (A6):\penalty0 4355--4364, 1981.

\bibitem[Peredo et~al.(1995)Peredo, Slavin, Mazur, and Curtis]{Peredo1995}
M.~Peredo, J.~Slavin, E.~Mazur, and S.~Curtis.
\newblock Three-dimensional position and shape of the bow shock and their
  variation with {Alfvenic}, sonic and magnetosonic {Mach} numbers and
  interplanetary magnetic field orientation.
\newblock \emph{Journal of Geophysical Research: Space Physics (1978–2012)},
  100\penalty0 (A5):\penalty0 7907--7916, 1995.

\bibitem[Petrukovich et~al.(2015)Petrukovich, Artemyev, Vasko, Nakamura, and
  Zelenyi]{Petrukovich2015}
A.~Petrukovich, A.~Artemyev, I.~Vasko, R.~Nakamura, and L.~Zelenyi.
\newblock Current sheets in the earth magnetotail: Plasma and magnetic field
  structure with cluster project observations.
\newblock \emph{Space Science Reviews}, 188\penalty0 (1-4):\penalty0 311--337,
  2015.

\bibitem[Pokhotelov et~al.(2013)Pokhotelov, von Alfthan, Kempf, Vainio,
  Koskinen, and Palmroth]{Pokhotelov2013}
D.~Pokhotelov, S.~von Alfthan, Y.~Kempf, R.~Vainio, H.~Koskinen, and
  M.~Palmroth.
\newblock Ion distributions upstream and downstream of the {Earth}'s bow shock:
  first results from {Vlasiator}.
\newblock In \emph{Annales {Geophysicae}}, volume~31, pages 2207--2212.
  Copernicus GmbH, 2013.

\bibitem[Pritchett(2000)]{Pritchett2000}
P.~L. Pritchett.
\newblock Particle-in-cell simulations of magnetosphere electrodynamics.
\newblock \emph{Plasma Science, IEEE Transactions on}, 28\penalty0
  (6):\penalty0 1976--1990, 2000.

\bibitem[Rojas-Castillo et~al.(2013)Rojas-Castillo, Blanco-Cano, Kajdič, and
  Omidi]{Rojas-Castillo2013}
D.~Rojas-Castillo, X.~Blanco-Cano, P.~Kajdič, and N.~Omidi.
\newblock Foreshock compressional boundaries observed by {Cluster}.
\newblock \emph{Journal of Geophysical Research: Space Physics}, 2013.

\bibitem[Samsonov(2007)]{Samsonov2007}
A.~A. Samsonov.
\newblock Specific features of magnetic barrier formation near the
  magnetopause.
\newblock \emph{Geomagnetism and Aeronomy}, 47:\penalty0 316--324, June 2007.
\newblock \doi{10.1134/S0016793207030061}.

\bibitem[Savoini et~al.(2013)Savoini, Lembege, and Stienlet]{Savoini2013}
P.~Savoini, B.~Lembege, and J.~Stienlet.
\newblock On the origin of the quasi-perpendicular ion foreshock:
  {Full}-particle simulations.
\newblock \emph{Journal of Geophysical Research (Space Physics)}, 118:\penalty0
  1132--1145, Mar. 2013.
\newblock \doi{10.1002/jgra.50158}.

\bibitem[Schreiner and Spanier(2014)]{Schreiner2014}
C.~Schreiner and F.~Spanier.
\newblock Wave-particle-interaction in kinetic plasmas.
\newblock \emph{Comput. Phys. Commun.}, 2014.

\bibitem[Seki et~al.(2009)Seki, Nishino, Fujimoto, Miyashita, Keika, Hasegawa,
  Okabe, Kasaba, Terasawa, Yamamoto, and {others}]{Seki2009}
Y.~Seki, M.~Nishino, M.~Fujimoto, Y.~Miyashita, K.~Keika, H.~Hasegawa,
  K.~Okabe, Y.~Kasaba, T.~Terasawa, T.~Yamamoto, and {others}.
\newblock Observations of loss cone–shaped back streaming energetic protons
  upstream of the {Earth}'s bow shock.
\newblock \emph{Journal of Geophysical Research: Space Physics (1978–2012)},
  114\penalty0 (A11), 2009.

\bibitem[Shaikhislamov et~al.(2011)Shaikhislamov, Antonov, Zakharov,
  Boyarintsev, Melekhov, Posukh, and Ponomarenko]{Shaikhislamov2011}
I.~Shaikhislamov, V.~Antonov, Y.~P. Zakharov, E.~Boyarintsev, A.~Melekhov,
  V.~Posukh, and A.~Ponomarenko.
\newblock Small scale magnetosphere: {Laboratory} experiment, physical model
  and {Hall} {MHD} simulation.
\newblock \emph{arXiv preprint arXiv:1110.4461}, 2011.

\bibitem[Treumann(2009)]{Treumann2009}
R.~Treumann.
\newblock Fundamentals of collisionless shocks for astrophysical application,
  1. {Non}-relativistic shocks.
\newblock \emph{The Astronomy and Astrophysics Review}, 17\penalty0
  (4):\penalty0 409--535, 2009.

\bibitem[Vapirev et~al.(2013)Vapirev, Lapenta, Divin, Markidis, Henri, Goldman,
  and Newman]{Vapirev2013}
A.~Vapirev, G.~Lapenta, A.~Divin, S.~Markidis, P.~Henri, M.~Goldman, and
  D.~Newman.
\newblock Formation of a transient front structure near reconnection point in
  3-{D} {PIC} simulations.
\newblock \emph{Journal of Geophysical Research: Space Physics}, 2013.

\bibitem[Villasenor and Buneman(1992)]{Villasenor1992}
J.~Villasenor and O.~Buneman.
\newblock Rigorous charge conservation for local electromagnetic field solvers.
\newblock \emph{Comput. Phys. Commun.}, 69\penalty0 (2):\penalty0 306--316,
  1992.

\bibitem[Welling et~al.(2013)Welling, Liemohn, Toth, and Glocer]{Welling2013}
D.~Welling, M.~Liemohn, G.~Toth, and A.~Glocer.
\newblock Evaluating the {Importance} of {Outflow} {Velocity} at the {MHD}
  {Inner} {Boundary}.
\newblock In \emph{{AGU} {Fall} {Meeting} {Abstracts}}, volume~1, page 2105,
  2013.

\bibitem[Winglee et~al.(2005)Winglee, Lewis, and Lu]{Winglee2005}
R.~Winglee, W.~Lewis, and G.~Lu.
\newblock Mapping of the heavy ion outflows as seen by {IMAGE} and multifluid
  global modeling for the 17 {April} 2002 storm.
\newblock \emph{Journal of Geophysical Research: Space Physics (1978–2012)},
  110\penalty0 (A12), 2005.

\bibitem[Wodnicka(2009)]{Wodnicka2009}
E.~Wodnicka.
\newblock On a magnetosphere disturbed by solar wind; observations of
  macroelectrons.
\newblock In \emph{Annales {Geophysicae}}, volume~27, pages 2331--2339.
  Copernicus GmbH, 2009.

\end{thebibliography}
     \clearpage
     \section{Tables}
\begin{table}[h]
	\centering 
	
	\begin{tabular}{||l||l||l||}
		\hline \hline
		\rowcolor{yellow}			 \parbox{3.0cm}{   Parameters} &  \textrm{PIC code} &   \textrm{GUMICS-v4}  \\ \hline
		\small  CPU time &  50 Min &  5 Hours \\   
		\small  Machine & \small  Single processor PC & \small CCMC-Super Computer \\ 
		\small      Ionosphere &  No &  Yes \\
		\small       \parbox{3.cm}{Grid Cell}  &  \small  Fixed & \small   \parbox{3cm}{Adaptive}  \\ 
		\small            \parbox{3.cm}{Grid size}  &  $ 1R_{E}^{3}  $&   \parbox{2.3cm}{$ (0.1-8R_{E})^{3} $}  \\  
		\small  Small Box Size & $ \small 155\times105\times105  R_{E} $   &  \small  $ 250\times130\times130 R_{E}   $\\
		\small        $\rho$& $ 0.8 \dfrac{N}{\Delta^{3}}  $&  $ 5.0 cm^{-3} $\\
		\small             $B_z(x)$& $ 0.2 $ &  $ 6.5 (nT) $\\
		\small              $V_{x}(x)$& $ 0.25  $&  500 $(km.sec^{-1}) $ \\
		
		\small            $ V_{A}$& $ 0.028 $ &   63$(km.sec^{-1} )$ \\
		&&\\
		
		&\parbox{4.cm}{\underline{\textbf{\boxed{Unitless \quad values}}}  }  &\\
		\hdashline 
		&&\\ 
		
		$\frac{V_{A}}{V_{SW}}$&$  0.11 $ & $ 0.12 $\\
		
		M$_A$ & 8.9 & 7.8\\
		
		$M_{MS}$& $ 5.5 $& $ 5.2 $\\
		$ \beta $&$ 1.6 $&$  2.7 $\\     			\hline \hline
	\end{tabular}\vskip0.5cm
	\caption{shows the solar wind input scaled parameters  for the  \textbf{PIC } and their corresponding values for the \textbf{ \textit{MHD}} code}\label{Tab:1}
\end{table} \clearpage
     \section{Figures}
     	% %
     	
     	\clearpage
     	% % % % % % % % Fig 2
     	
     	% % %
     	\newpage
     	% % % % % % % % Fig 2

     	%%%%%
     	% %
     	
     	% fig 2
     	
     	% ---------------
     	% % % % % %
     	
     	% % % %
     	% %
     	
     	% EXAMPLE TABLE
     	
     	
     \end{document}